\documentclass[useAMS,usenatbib,a4paper]{mn2e}
\usepackage{amssymb,amsmath}
\usepackage{graphicx}
\usepackage{natbib}
\usepackage{journal_shortcuts}
\newcommand{\Sun}{_{\sun}}

\newcommand{\ICM}{_{\mathrm{ICM}}}

\newcommand{\K}{\,\textrm{K}}
\newcommand{\Kpc}{\,\textrm{kpc}}

\newcommand{\Myr}{\,\textrm{Myr}}

\newcommand{\Kms}{\,\textrm{km}\,\textrm{s}^{-1}}

\newcommand{\gccm}{\,\textrm{g}\,\textrm{cm}^{-3}}

\title[Ram pressure stripping in a viscous ICM]%
{Ram pressure stripping in a viscous intracluster medium}

\author[E. Roediger and M. Br\"uggen]%
{Elke Roediger%
\thanks{E-mail:
e.roediger@jacobs-university.de; m.brueggen@jacobs-university.de}
and  
Marcus Br\"uggen\footnotemark[1]%
\\
Jacobs University Bremen, P.O. Box 750\,561, 28725 Bremen,
Germany}

\begin{document}

\date{Accepted. Received; in original form }

\pagerange{\pageref{firstpage}--\pageref{lastpage}} \pubyear{2008}

\maketitle

\label{firstpage}

\begin{abstract}

In the recent literature there is circumstantial evidence that the
viscosity of the intracluster medium may not be too far from the
Spitzer value. In this letter, we present two-dimensional
hydrodynamical simulations of ram pressure stripping of disc
galaxies in a viscous intracluster medium (ICM). The values of
viscosity explored range between 0.1 and 1.0 times the Spitzer
value. We find that viscosity affects the appearance and the
dimensions of the galactic wakes but has very little effect on the
evolution of the gas mass of the galaxy.

\end{abstract}


%
%
%
%
\section{Introduction}
%
In clusters, galaxies can lose some or all of their gas by ram pressure
stripping (RPS) due to their motion through the intracluster medium
(ICM). Both, analytical estimates (\citealt{gunn72}) and (hydro)dynamical
simulations (e.g.~\citealt{abadi99}, \citealt{quilis01}, \citealt{schulz01},
\citealt{vollmer01}, \citealt{marcolini03}, \citealt{roediger06},
\citealt{roediger07}) show that RPS can remove a significant amount of gas
from galaxies, and is thus important for the evolution of galaxies and the
ICM.  

In addition to gas loss by ram pressure pushing, galaxies also suffer gas loss
by continuous (sometimes also called turbulent or viscous) stripping
(e.g.~\citealt{nulsen82,quilis00,schulz01,roediger05,roediger06}).

\citet{nulsen82} has studied the effects of transport processes and turbulence
on the flow of gas past a galaxy and has found that they could produce more
stripping of gas than ram pressure alone.  For turbulent stripping, he found a
mass stripping rate of
\begin{equation}
\dot{M}_{\rm turb}\sim \pi r^2 \rho_{\rm ICM} v \sim 7\, r_{10}^2
n_{-3}v_3 M_{\odot} {\rm yr}^{-1} ,
\end{equation}
where $r=10r_{10}$ kpc, $\rho_{\rm ICM}=10^{-3}n_{-3}m_{\rm H}$ g cm$^{-3}$
and $v=1000 v_3$ km s$^{-1}$. This is similar to what is found in the
simulations by \citet{roediger06, roediger07} who find rates of about $\sim 1
M_{\odot} {\rm yr}^{-1}$.  Turbulent stripping is mainly caused by the
Kelvin-Helmholtz instability, which is suppressed by viscosity that stabilises
modes with wavelengths smaller than $r/{\rm Re}$ (e.g. \citealt{betchov67}).

There exist only few detailed studies on the wakes of
stripped galaxies. Using the hydrodynamical adaptive mesh refinement code
FLASH, we have studied galactic ram pressure tails in a constant ICM wind
(\citealt{roediger06wakes}) as well as for galaxies on realistic cluster
orbits (\citealt{roediger08}). The minimum tail width of about 20 to 30 kpc is
found near the galaxy. With increasing distance to the galaxy, the tail flares
to widths of 30 to 80 kpc at a distance of $\sim 100\Kpc$ behind the
galaxy. Other hydrodynamical simulations using either grid codes
(e.g.~\citealt{quilis01}, \citealt{marcolini03}) or smoothed particle
hydrodynamics (e.g.~\citealt{abadi99}, \citealt{schulz01}) do not focus on the
tails. However, according to the snapshots provided in these papers, the gas
tails are similar to the ones in our simulations.  Using a sticky-particle
code, Vollmer et al.~(e.g.~1999, 2000, 2001a, 2001b, 2003, 2004a, 2005,
2006\nocite{vollmer99,vollmer00,
vollmer01,vollmer01a,vollmer03,vollmer04,vollmer05,vollmer06}) aim at
reproducing the RPS history of individual galaxies by comparing simulations
and observations. However, these simulations concentrate on the gas
distribution close to the galaxy.

Recently, \citet{oosterloo05} presented observations of a $\sim 120$ kpc long
and $<25\Kpc$ wide tail of H~I gas associated with NGC~4388, and also
suggest that this tail is due to ram-pressure stripping of this galaxy, either
in the ICM of the Virgo cluster or in the halo of the nearby elliptical galaxy
M86. However, the tail of NGC~4388 is narrower than the tails in the
simulations presented in \citet{roediger06, roediger08}, although it seems to
be flaring in a similar fashion as found in the simulation. Also the X-ray and
H$\alpha$ tails observed by \citet{sun06,sun07,yagi07} are much narrower than
simulated ram pressure tails: While being only $\sim 7\Kpc$ wide, they
reach lengths of $\sim 70\Kpc$ and show hardly any flaring. This difference
may be caused by the microphysics of the ICM.

The viscosity of the intracluster medium has been discussed before. Based on
observations of the Perseus cluster it has been suggested by \cite{fabian03a},
\cite{fabian03b} that viscosity may play an important role in dissipating
energy injected by the central AGN.  Circumstantial evidence for the presence
of significant ICM viscosity is also provided by an examination of the
morphology of H$\alpha$ filaments in the Perseus cluster. Several of the
filaments appear to trace well-defined arcs which argues against the presence
of strong turbulence in the ICM core, possibly resulting from the action of
viscosity. This idea has been tested in numerical simulations by
\citet{ruszkowski04, ruszkowski04a} and \citet{reynolds05}.

In the case of a fully ionized and unmagnetised, thermal plasma, the relevant
coefficient of viscosity is given by \citet{braginskii58} and \cite{spitzer62}
as $\mu \approx 6.0\times 10^{-17}(\ln\Lambda/37)^{-1}T^{5/2}$ g cm$^{-1}$
s$^{-1}$, where $T$ is the temperature of the plasma measured in Kelvin and
$\ln \Lambda$ is the Coulomb logarithm. It results from the cumulative
effect of weak Coulomb collisions. The mean free path of such interactions
scales with $T^2$. As the sound speed scales with $T^{1/2}$, the viscosity is
proportional to $T^{5/2}$.

It is customary to measure the importance of viscosity through the Reynolds
number, Re $=vl/\nu$, where $v$ and $l$ are characteristic velocities and
length-scales of the system and $\nu=\mu/\rho$ the kinematic viscosity, with
$\rho$ being the fluid density.  For our case

\begin{equation}
{\rm Re} \sim 26\, v_3 r_{10}n_{-3}f_v^{-1}\left (\frac{kT}{5
{\rm keV}} \right )^{-2.5} , 
\end{equation}
which indicates that viscosity may play a role unless the viscosity is
strongly suppressed, i.e. the suppression factor, $f_{v}$,
is sufficiently small.

This suppression results from the cluster magnetic fields. Even weak fields
lead to a tiny proton gyroradius, which results in a very efficient
suppression (factor of $\sim 10^{23}$ for typical ICM conditions,
\citealt{spitzer62}) of the local viscosity perpendicular to the magnetic
field.  Magnetic fields in the ICM are certainly tangled or even chaotic
(\citealt{clarke04,ensslin05}) and will lead to a reduced macroscopic
viscosity. The degree of reduction, however, is unknown. Due to the same
mechanism, also thermal conduction in the ICM is suppressed perpendicular to
the magnetic field lines. In a recent study, \citet{narayan01} found an
effective macroscopic thermal conductivity that is a factor of $f_{v} \sim
10^{-2}-0.2$ lower than the unmagnetised value. Similar arguments may apply to
the viscosity. From studies of ICM turbulence in the Coma cluster,
  \citet{schuecker04} derive an upper limit on the kinematic viscosity of the
  ICM of $\sim 3\cdot 10^{29} \mathrm{cm}^2 \mathrm{s}^{-1}$. For typical ICM
  densities and temperatures, this corresponds to a viscosity suppression
  factor $f_{v}$ around 0.1. Thus, we might expect the viscosity to be
  suppressed by some factor between $10^{-2}$ and unity.

The magnetic fields of the ICM (e.g.~\citealt{clarke04,ensslin05}) may
themselves influence the appearance of ram pressure stripped galaxies. Given
that the thermal pressure in the ICM dominates over the magnetic pressure, the
magnetic fields should be frozen-in and follow the ICM flow. Thus, they should
be generally parallel to the galaxy's tail. Such a magnetic field structure
could attenuate thermal conduction between the stripped gas and the hot ICM,
and it could suppress e.g.~Kelvin-Helmholtz and maybe even Rayley-Taylor
instabilities in the tail. However, it is unclear how well the magnetic fields
will be aligned with the ICM-ISM interfaces and how strong this suppression
will be. The influence of the magnetic fields will be studied in a forthcoming
paper.

Here, using adaptive-mesh, hydrodynamical simulations, we investigate
the effect of a macroscopic viscosity on the stripping of gas from a
galaxy.

\section{Method}
%
We model the ICM-ISM interaction in the galaxy's rest frame, i.e. the
galaxy is exposed to an ICM flow. The work of \citet{roediger07}
showed that the classical estimate of the stripping radius given by
\citet{gunn72} can be adapted to galaxies exposed to a variable ICM
wind.  Therefore, here we decouple the effect of time-variability and
use a constant ICM wind to focus on the effect of viscosity.

To include viscosity in our hydro-simulations, we add velocity diffusion to
the momentum equation
\begin{equation}
\frac{\partial (\rho v_{i})}{\partial t}+
\frac{\partial}{\partial x_{k}}(\rho v_{k}v_{i})+
\frac{\partial P}{\partial x_{i}} = \rho g_{i}+
\frac{\partial\pi_{ik}}{\partial x_{k}} ,
\end{equation}
with
\begin{equation}
\pi_{ik}=\frac{\partial}{\partial x_{k}}\left[2\mu
\left(e_{ik}-\frac{1}{3}\Delta\delta_{ik}\right)\right],
\end{equation}
where $P$ is pressure, $\rho$ density, $v_i$ the components of  
velocity, $g_i$ the components of the gravitational acceleration, $ 
\delta_{ik}$ the Kronecker delta,
\begin{equation}
e_{ik}= \frac{1}{2} \left (\frac{\partial u_i}{\partial x_k} +\frac 
{\partial u_k}{\partial x_i}\right ),
\end{equation}
and $\Delta = e_{ii}$.
Here $\mu$ is the coefficient of (shear) viscosity. In our  
simulations we neglected bulk viscosity since it vanishes for an  
ideal gas.
 We use the
standard Spitzer viscosity for an unmagnetised plasma, for which $\mu
= 6.0\times 10^{-17}(\ln\Lambda/37)^{-1}T^{5/2}f_{v}$ g cm$^{-1}$
s$^{-1}$.

We have noted earlier that the precise value of the suppression factor
is highly uncertain and, depending on the nature of magnetic
turbulence, may even exceed the Spitzer value (\cite{cho03}) or be
suppressed well below it.  For our viscous simulations we choose
$f_{v}$ between $0$ and 1 (see Table~\ref{tab:runs}). 

In order to ensure numerical stability and to prevent too small timesteps we
have switched off viscosity for temperatures outside the range $3\cdot 10^5\K
< T < 1.5\cdot 10^8\K$.

The total energy fluxes are not modified by viscosity as the effects
of viscous heating are small.

\subsection{Code} \label{sec:code}
The simulations were performed with the FLASH code (\citealt{fryxell00})
version 2.5, a multidimensional adaptive mesh refinement hydrodynamics code.
It solves the Riemann problem on a Cartesian grid using the
Piecewise-Parabolic Method (PPM).

The viscous runs are fairly expensive because the timestep imposed by
the viscosity scales as $(\Delta x)^{2}/\mu$, where $\Delta x$ is the
resolution of the computational grid. As this timestep scales more
strongly with $\Delta x$ than the standard hydrodynamical Courant
condition, and because viscosity depends strongly on temperature, the
constraints on the timestep are more stringent than in the inviscid
runs. Hence, here we present only results from 2D simulations in cylindrical
coordinates, $(R,Z)$. 

The ICM wind enters the simulation box at $Z=-65\Kpc$. The boundary at $R=0$ is the
galaxy's symmetry axis, the remaining two boundaries obey outflow conditions.
We use a grid size of $260\Kpc$ in $Z$-direction $65\Kpc$ in $R$-direction.
With 5 refinement levels we reach an effective number of 2048 $\times$ 256
grid cells and effective resolution of 0.25 $\Kpc$.

\subsection{Model galaxy}
The galaxy model is the same as in \citet{roediger06}, i.e. a
massive spiral with a flat rotation curve at $200\Kms$. It consists of a dark
matter halo ($1.1\cdot 10^{11}M\Sun$ within $23\Kpc$), a stellar bulge
($10^{10}M\Sun$), a stellar disc ($10^{11}M\Sun$) and a gaseous disc ($5\cdot
10^{9}M\Sun$). All non-gaseous components just provide the galaxy's potential
and are not evolved during the simulation. For a description of the individual
components and a list of parameters please refer to RB06\nocite{roediger06}.
Initially, the gas disc is set in hydrostatic equilibrium with the surrounding
ICM (see also RB06). The disc's rotation is included via the
  centrifugal force.

\subsection{ICM conditions}
Table~\ref{tab:runs} lists the ICM conditions for the simulation runs. The
ICM temperature was $7.2\cdot 10^7$ K or 6.2 keV in all runs.
%
\begin{table}
\caption{ICM conditions for different simulation runs.}
\label{tab:runs}
\centering\begin{tabular}{ccccc}
\hline
  run:                   & M08 & M08-V1 & M08-V2 & M08-V3 \\
\hline				                      
$\rho\ICM /10^{-27}\gccm$ &  1  &   1  &   1   &   1      \\
$v\ICM /1000\Kms$        & 0.8 &  0.8 &  0.8  &  0.8    \\
$f_{v}$                  &  0  &  0.1 &  0.5  &  1.0     \\
\hline
\end{tabular}
\centering\begin{tabular}{ccc}
\hline
  run:                   & M20 & M20-V \\
\hline				                      
$\rho\ICM /10^{-27}\gccm$ & 1    &   1    \\
$v\ICM /1000\Kms$        &  2  &  2     \\
$f_{v}$                  &  0  & 0.1   \\
\hline
\end{tabular}
\end{table} 
%
We start the simulation with the ICM at rest and then increase the inflow
velocity over the first $50\Myr$ from zero to $v\ICM$ (for more details see
RB06).

\section{Results}
%
\begin{figure*}
\centering\resizebox{0.99\hsize}{!}{\includegraphics{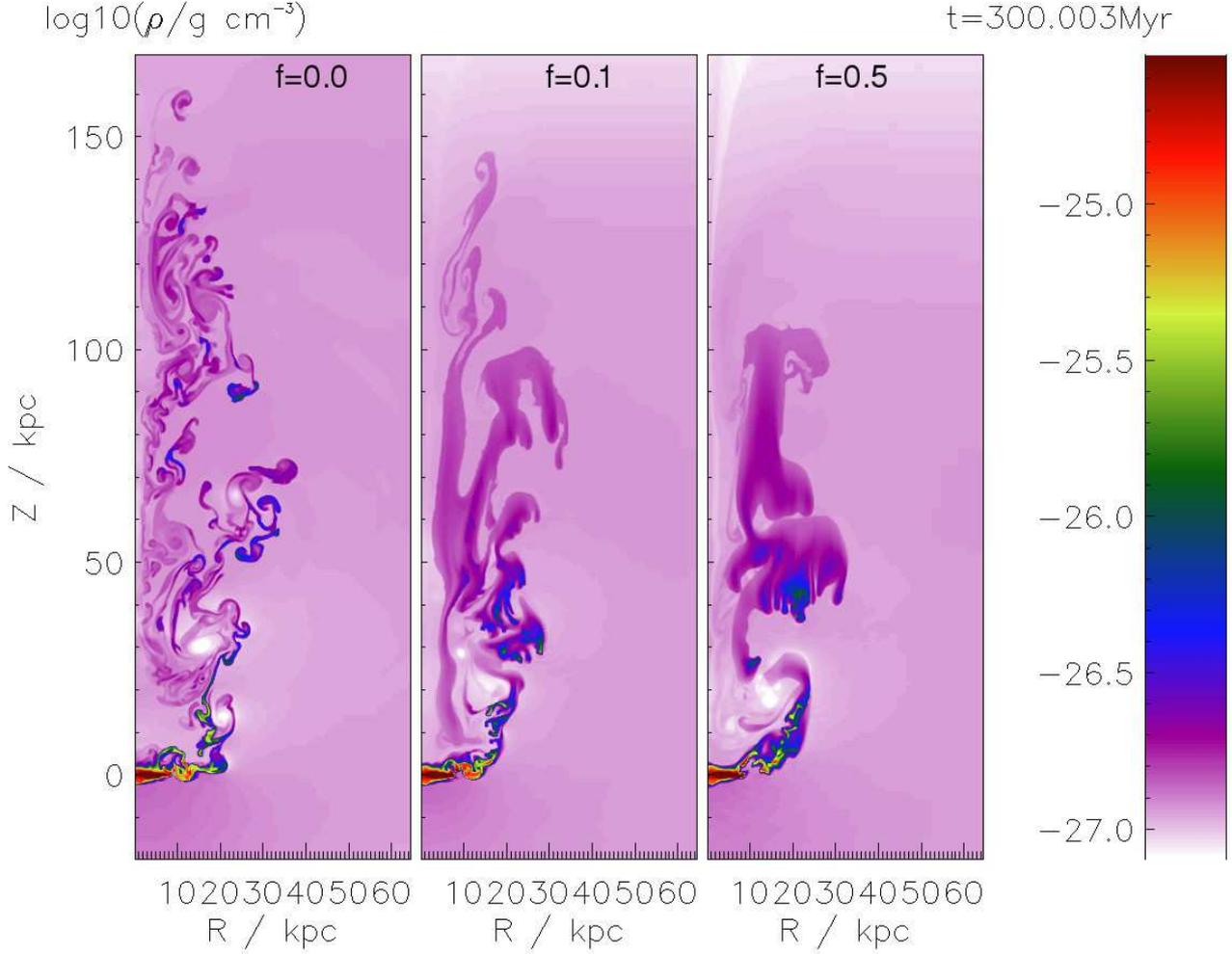}}
\caption{Contour plot of the density in subsonic runs at 300 Myr, for
  increasing viscosity from left to right: left is inviscid ($f_{\rm v}=0$),
  middle is viscous with $f_{\rm v}=0.1$, right is viscous with $f_{\rm
  v}=0.5$.}
\label{fig:density}
\end{figure*} 
%
Figure~\ref{fig:density} shows the density in
the subsonic runs with increasing viscosity: $f_v=0, 0.1$ and 0.5. The
different behaviour of the galactic wakes is evident: The larger the
viscosity, the less turbulent the wake.  In the viscous cases, the ICM
flows past the galaxy more smoothly than in the non-viscous runs. In
the non-viscous cases, the KH- and RT-instability cause vortices and
turbulence. Both instablilities are suppressed significantly in the
viscous cases. Consequently, the stripped gas remains in larger
clumps and is less readily mixed with the ambient medium.

Also the width of the tail, or the flaring angle, is slightly smaller in the
viscous runs. In the inviscid run, the maximum width of the tail is $\approx
2\times 40$ kpc, while it is $\approx 2\times 35$ kpc in all viscous
runs. Thus, it is also unlikely that the small width of the tail in NGC~4388
and the X-ray tails of ESO137-001 (\citealt{sun06}) can be explained by a
microphysical, Spitzer-type viscosity. Interestingly, the tail appears to be
shorter, the larger the viscosity. This is due to the fact that larger gas
clouds, even if they have the same density like smaller clouds, experience a
smaller acceleration for a fixed ram pressure. We note that the mass
fraction of stripped gas in dense form does not depend strongly on
viscosity. In the viscous cases, there are a few large dense clouds, whereas
in the non-viscous case there are numerous small dense clouds.  We do not
show any plots here for the supersonic runs but they show the same qualitative
behaviour.

\begin{figure}
\centering\resizebox{0.99\hsize}{!}{%
\includegraphics{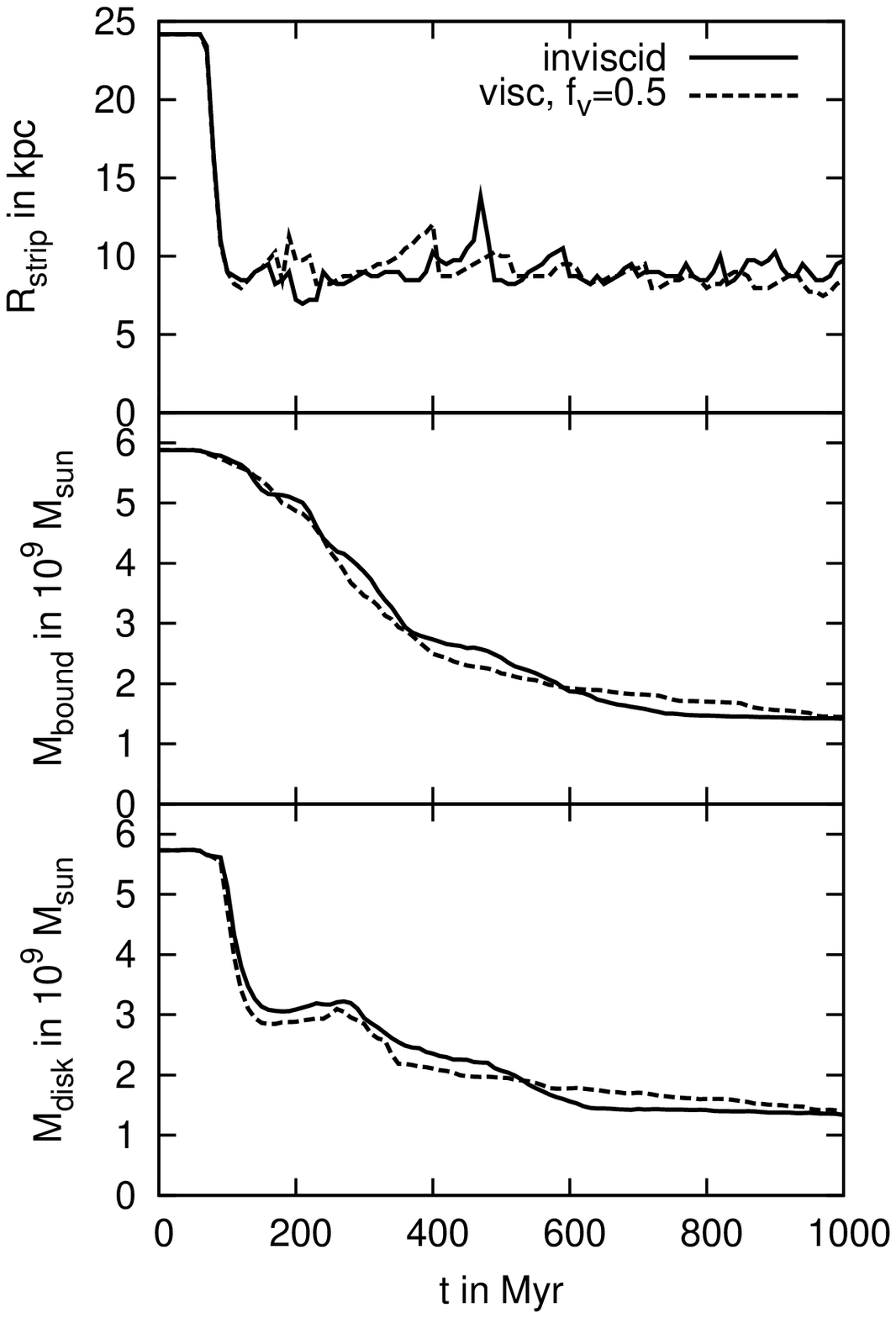}}
\caption{Evolution of radius (top panel) and mass of the remaining gas disc:
  the bottom panel measures the
  mass in a fixed cylinder centred on the galaxy, the middle panel measures
  the mass gravitationally bound to the galaxy. The viscous runs are for
  $f_{v}=0.5$.}
\label{fig:mass_radius}
\end{figure}

Figure~\ref{fig:mass_radius} compares the evolution of the mass and radius of
the remaining gas disc for viscous and inviscid runs.  The top panel
demonstrates that the viscosity has nearly no influence on the evolution of
the gas disc's radius. 
 The bottom panel displays the gas mass in a cylinder
of radius 27~kpc and a height of 10~kpc. Interestingly, the mass evolution in
all runs is very similar, for all values of the viscosity that we have
explored. The first dip in the graph marks the end of the instantaneous
stripping phase. Very slight differences introduced by the viscosity occur
during the next, the intermediate phase and also during the final continuous
stripping phase.  We have also measured the gas mass bound to the
gravitational potential of the galaxy and its behaviour is shown in the middle
panel of Fig.~\ref{fig:mass_radius}. The differences between runs with and
without viscosity is similar to that of the gas disc mass. This implies that
the amount of gas lost from the galaxy is fixed by the ram pressure.

This result shows that the viscosity in the parameter range considered here
has a minor impact on the mass loss history of the gas disc and that ram
pressure pushing is the dominant mechanism, as found in earlier papers
(e.g.~\citealt{roediger06,roediger07}). The fact that the viscosity has a
minor impact on the mass loss history should also hold for most galaxies which
move though clusters and thus experience a variable ram pressure -- for the very
reason that ram pressure pushing is the main cause for mass loss also for
these galaxies (\citealt{roediger07}). The
mass loss due to ram pressure pushing, i.e.~during the first,
instantaneous stripping phase, is especially insensitive to different ICM viscosities.
An exception may occur when galaxies
move near edge-on.
%
%
\section{Discussion}
%
Clearly, we have to be careful with quantitative predictions
especially concerning the morphology of the wake because our
simulations are two-dimensional. However, in previous work we have
shown that mass loss rates are the same in 2D and 3D
simulations. Moreover, in the non-viscous cases, the wake structures
in 2D and 3D were very similar. Thus, our main conclusions should be
robust:
\begin{itemize}
\item The mass loss from the gas discs is hardly influenced by viscosity.
\item With increasing viscosity, the wake shows less stucture and turbulence,
  but larger clumps.
\end{itemize}

A detailed investigation of the fate of the stripped gas requires
several additions to our simplified model: a 3D treatment to allow for
other inclinations than face-on, prescriptions for heating, cooling
and thermal conduction. These processes do not only influence the
temperature in the stripped gas, but also determine in which wavebands
the stripped gas will be observable. For the fate of the stripped
gas, viscosity may make a difference as it may affect the thermal
history of the stripped gas. Gradients of density and temperature are
smeared out in the wake in the presence of viscosity. This will affect
the efficiency of heat conduction and evaporation of clumps in the
galactic wakes. It will also affect the efficiency with which material
in the wake can form stars as observed by \citet{sun07}. Comparisons
between models and observations will reveal which processes are the dominant
ones in shaping galactic tails.


\section*{Acknowledgements}
We acknowledge the support by the DFG grant BR 2026/3 within the Priority
Programme ``Witnesses of Cosmic History'' and the supercomputing grants NIC
2195 and 2256 at the John-Neumann Institut at the Forschungszentrum J\"ulich.
The results presented were produced using the FLASH code, a product of the DOE
ASC/Alliances-funded Center for Astrophysical Thermonuclear Flashes at the
University of Chicago. We thank the referee for the helpful comments.


%
\bibliographystyle{mn2e}
\bibliography{%
BIBLIOGRAPHY/theory_simulations,%
BIBLIOGRAPHY/hydro_processes,%
BIBLIOGRAPHY/numerics,%
BIBLIOGRAPHY/observations_general,%
BIBLIOGRAPHY/observations_clusters,%
BIBLIOGRAPHY/observations_galaxies,%
BIBLIOGRAPHY/galaxy_model,%
BIBLIOGRAPHY/gas_halo,%
BIBLIOGRAPHY/icm_conditions,%
BIBLIOGRAPHY/clusters,%
BIBLIOGRAPHY/agn,%
BIBLIOGRAPHY/else}

\bsp

\label{lastpage}

\end{document}